\documentstyle[12pt]{article}
\newcommand{\beq}{\begin{equation}}
\newcommand{\eeq}{\end{equation}}

\newcommand{\eq}[1]{eq.($\ref{#1}$)}

\title{\bf Nuclear Structure-Dependent\\ Radiative Corrections \\
to the Hydrogen Hyperfine Splitting\\}
\author{
Savely G. Karshenboim\thanks{ E-mail: sgk@onti.vniim.spb.su;
karshenboim@phim.niif.spb.su}\\
\bigskip
\sl D. I.  Mendeleev Institute for Metrology (VNIIM),\\
\sl St. Petersburg 198005, Russia}

\date{}

\begin{document}


\maketitle

\bigskip

\bigskip


\bigskip

\bigskip

KEY words: hyperfine structure, radiative correction, proton polarizability

\newpage

\begin{abstract}

Radiative corrections to the Zemach contribution of the
hydrogen hyperfine splitting are calculated. Their contributions amount 
to 
$-0.63(3)\,ppm$ to the HFS. The radiative recoil corrections are estimated to be
$0.09(3)\,ppm$ and heavy particle vacuum polarization shifts the HFS by 
$0.10(2)\,ppm$. The  status of the
nuclear-dependent contributions are considered. From the comparison of 
theory and experiment
the proton polarizability contribution of $3.5(9)\,ppm$ is found.
The nuclear structure-dependent corrections
to the difference $\nu_{hfs}(1s) -n^3\nu_{hfs}(ns)$ are also obtained. 

\end{abstract}

\newpage


\bigskip

\section{Introduction}

The hyperfine splitting of the ground state of the hydrogen atom is one 
of the most
precise measured values \cite{HFS1,HFS2}

\beq
\nu_{HFS}(1s) = 1420405.7517667(9)~kHz,
\eeq

\noindent
but the theory is not able to obtain this result with such an accuracy. 
The main problem
is due to the proton structure. This work is devoted to nuclear 
structure-dependent
contributions.

An expression of the leading nuclear structure-dependent correction to 
the hyperfine splitting in the hydrogen atom was found by Zemach \cite{Zem}.
Later nuclear
structure-dependent corrections were investigated in Refs.
\cite{Arn,New,Idd,Gro,Bod}, but the radiative structure-dependent
corrections have not yet been found. The pure radiative corrections are 
evaluated
in the next section. The third section is devoted to radiative recoil 
corrections.
After that  the vacuum polarization of heavy particle 
is discussed. 
The last sections 
consider the status of the nuclear 
structure-dependent terms.


\section{External Field Approximation}

\subsection{Leading structure-dependent term}

The Zemach expression of the leading nuclear structure-dependent correction 
to the ground state hyperfine splitting in the hydrogen atom has the form 
\cite{Zem}

\[
\Delta \nu(Zemach) = \nu_F \left(-2Z\alpha m_e
\int
{\int{d^3{\bf r}}\,
d^3{\bf r}'}
\rho_E({\bf r})\rho_M({\bf r'})|{\bf r}-{\bf
r'}|\right),
\]

\noindent
or

\beq \label{ZemEx}
\Delta \nu(Zemach) = \nu_F\, \frac{2Z\alpha m_e}{\pi^2}
\int{\frac{d^3{\bf p}}
{{\bf p}^4}}\left[\frac{G_E({\bf p}^2)G_M({\bf p}^2)}{1+\kappa}-1\right]
\eeq

\noindent
where $\rho_{E}({\bf r})$ and $\rho_{M}({\bf r})$ are the proton
electric charge and magnetic moment distribution respectively, 
$G_{E/M}({\bf p}^2)$ is the Sachs
electric/magnetic form factor, $\kappa$ is the proton anomalous 
magnetic moment. The
Fermi energy $\nu_F$ is defined as the ground state hyperfine splitting in the nonrelativistic 
theory. Here, relativistic units 
in which
$ \hbar  = c = 1$ and $ \alpha=e^2 $ are used. $Z$ is the nuclear charge 
in units of 
the proton charge. It is equal to one in the
hydrogen atom, but some results in this paper like \eq{ZemEx} are valid
for any low-$Z$ hydrogen-like atom.

\subsection{Dipole approximation}

In the well-known dipole approximation of the electromagnetic form factors 
of the proton

\beq  \label{Dipo}
G_E({\bf p}^2)=\frac{G_M({\bf p}^2)}{1+\kappa}
= \left[\frac{\Lambda^2}{\Lambda^2+{\bf p}^2}\right]^2
\eeq

\noindent
it is easy to solve \eq{ZemEx} \cite{Bod}:

\beq \label{ZemEx1}
\Delta \nu(Zemach) = \nu_F \left(-\frac{35}{4}
\frac{Z\alpha m_e}{\Lambda}\right).
\eeq

\subsection{Vacuum polarization}

To calculate the contribution of the electronic vacuum polarization
one can insert the well-known asymptotic behaviour of the polarization
operator

\beq \label{VP-e}
2 \, \frac{\alpha}{\pi} \,
\left\{\frac{1}{3}\log{\frac{{\bf p}^2}{m^2_e}}-\frac{5}{9}\right\}
\eeq

\noindent 
into the right hand side of \eq{ZemEx}.
Integrating \eq{ZemEx} within the dipole approximation(\eq{Dipo})
yields:

\[
\Delta \nu(structure-VP) = \Delta \nu(Zemach)\cdot
\frac{\alpha}{\pi} \,
\left\{\frac{2}{3}\log{\frac{\Lambda^2}{m^2_e}}-\frac{634}{315}\right\}
.\]

\noindent This result is in agreement with an estimate found in Ref.
\cite{Bod}.

The logarithmic part of this result may be used for 
any low-$Z$ hydrogen-like atom
(if the magnetic square radius ($R_M$) is aproximately equal 
to the charge 
square radius ($R_E$)) after substituting $\Lambda \to \sqrt{12}/R_E$.

\subsection{Self energy}

The self-energy contribution is evaluated by using an explicit
asymptotic expression of the one-loop insertion into the electron line. 
The result is

\beq  \label{Str-SE}
\Delta \nu(structure-{\mbox{\it e-line}}) = \Delta \nu(Zemach) \cdot
\frac{\alpha}{\pi} \, \left\{-\frac{5}{4}\right\}
.\eeq

\noindent 
The coefficient (-5/4)  arises  from nontrivial  radiative insertion into 
the electron line  (-7/4) \cite{KSE} and from the anomalous magnetic moment 
contribution  (1/2). 
It should be mentioned that this result has been obtained without
the use of the  dipole fit (\eq{Dipo}) and it is also valid for any low-$Z$
hydrogen-like  atom. It can be used in a wide interval of the nuclear charge
up to 
$Z=25$ and the uncertainty is expected as to grow $(\pi Z \alpha)^2$ in
relative units.

\subsection{Binding corrections}

The higher order binding corrections have the order $(Z\alpha)^2$ or 
$(Z\alpha\,m_e)/\Lambda$ in units of $\Delta\nu(Zemach)$. They are small and 
it is enough to take into account only their logarithmic parts. The 
$(Z\alpha)^2$-term arises from the Dirac correction to the wave function 

\[
\Delta \nu(structure-Dirac) = \Delta \nu(Zemach)\cdot\frac{(Z\alpha)^2}{2}
\log{\frac{1}{(Z\alpha)^2}}.
\]

\noindent The $(Z\alpha\,m_e)/\Lambda$-contribution {\sc is due to} the 
nuclear charge distribution and the Fermi interaction. The result is

\[
\Delta \nu(structure-charge) = \Delta \nu_F \cdot
\left( - \frac{2}{3} (Z\alpha m_e)^2  r_p^2 
\log{\frac{1}{(Z\alpha)^2}} \right)
.\]

These corrections are small, but they are considered here, because only 
the binding corrections can 
contribute to the hfs splitting of the higher-$l$  state or to the 
difference

\[
\Delta_{hfs}(n)= \nu_{hfs}(1s) -n^3\nu_{hfs}(ns).
\]

The result for the difference can be found in the non-relativistic 
approximation (cf. Ref.
\cite{IK96NP})

\beq  \label{Bi1}
\delta\Delta_{hfs}(n) = \Delta \nu(Zemach) \cdot (Z\alpha)^2
\left[ \psi(n+1)-\psi(2) - \log{n} - \frac{(n-1)(n+9)}{4n^2} \right],
\eeq

\noindent where $\psi(z) = (d/dz)\log\Gamma(z)$, and for the hfs of states 
with $l>0$

\[
\delta \nu(nl_j) = \frac{\Delta \nu(Zemach)}{n^3} \cdot (Z\alpha)^2\,
\frac{n^2-1}{4n^2}\, \delta_{j,1/2}\delta_{l1}.
\]

\noindent The $(Z\alpha\,m_e)/\Lambda$-contribution may be found 
in the same technics useds in Ref. \cite{log2}

\beq  \label{Bi2}
\delta\Delta_{hfs}(n) =
  \Delta \nu_F \cdot
\left( - \frac{2}{3} (Z\alpha m_e)^2  r_p^2 \right)\,
\left(\frac{n-1}{n}-\log{n}+\psi(n)-\psi(1)\right)
\eeq

\noindent The binding corrections of \eq{Bi1} and \eq{Bi2} shift 
the difference $\Delta_{hfs}(2)$ by $10^{-9}\,$kHz only.


\section{Recoil Contributions}

\subsection{The leading term}

The final result of the pure recoil corrections was found in Ref.
\cite{Bod} by  numerical means  using 
the dipole approximation. It includes large numerical 
cancelation between different terms. This cancelation can be understood 
analytically from the leading logarithmic term \cite{Arn,New} 
for atoms with a non-structured nucleus (terms $VO$, $VV$ and $\kappa^2$ 
of Ref. \cite{Bod})

\beq   \label{ArnR}
\Delta \nu(rec) = - \nu_F \cdot
\frac{3Z\alpha}{\pi}\frac{m_e}{m_p}\log{\frac{m_p}{m_e}}\cdot
\frac{2(1+\kappa)-(1+\kappa)^2+\frac{3}{4}\kappa^2}{1+\kappa}.
\eeq

\noindent 
The coefficient 

\[
\frac{2(1+\kappa)-(1+\kappa)^2+\frac{3}{4}\kappa^2}{1+\kappa} 
\]

\noindent
is equal to 1 in muonium and $0.070...$ in the hydrogen atom.\\

\smallskip

In the following subsections radiative recoil corrections (i. e. 
radiative corrections to 
\eq{ArnR}) are considered. 

\subsection{Electronic vacuum polarization}

The electronic vacuum polarization term contains the same structure of the 
Dirac matrix as the leading term  of \eq{ArnR} and the same cancelation 
occurs. The estimate 

\[
\Delta \nu(VP-log) = - \nu_F \cdot
\frac{2\alpha(Z\alpha)}{\pi^2}\frac{m_e}{m_p}\log^2{\frac{\Lambda}{2m_e}}
\]

\[
\times
\frac{2(1+\kappa)-(1+\kappa)^2+\frac{3}{4}\kappa^2}{1+\kappa}.
\]

\noindent
can be obtained by using the asymptotics of \eq{VP-e} (see also \cite{Bod}).

\subsection{Self energy}

 The self-energy recoil contribution can be estimated from this 
contribution in the muonium atom, which includes a non-relativistic pole 
("the $\delta'$-term" in the definitions used in Ref. \cite{STY}  or 
"the $NR$-contribution" in Refs. \cite{EKS91,BEKS}), logarithmic and
constant  contributions of two different structures of the Dirac matrix
\cite{STY,EKS91,BEKS}. The numerically important contributions arise from
the pole and logarithmic terms. They can easily be adjusted to the hydrogen
atom. The constant can be used to estimate unsertainty. The  anomalous
magnetic moment contribution has to be added as well. The final  numerical
result is presented in the last section.

\section{Heavy Particle Vacuum Polarization}
 
The correction due to muonic and hadronic vacuum polarization have
been treated for a point-like nucleus in Ref. \cite{JP95}\footnote{That work 
contains some misprints: the left part of eq. (9) should be multiplied by 
$2\pi$; the result for the hadronic contribution to the HFS in eq.(10) 
and in the Table should be multipied by 2.}. 
These corrections are found here  in the 
external field approximation (for details see Ref. \cite{JP95}) with 
the dipole 
form factors of \eq{Dipo}. The numerical results are presented in 
the last section. The results for the point-like nucleus 

\[
\Delta \nu ( 
\mu-VP-point)=\frac{3}{4}\,\alpha(Z\alpha)\,\frac{m_e}{m_\mu}\, \nu_F
,
\]

\noindent and 

\[
\Delta \nu (hadr-VP-point) \simeq 0.7(3)\, \Delta \nu(\mu-VP-point)
\]

\noindent and for the  finite-size nucleus are different. The
finite-size  nucleus results are only some the  30\% of the point-like 
nuclear corrections.
The results for the finite-size nucleus are given in this work 
within the external field
approximation. The uncertainties are estimated by the unknown recoil
contributions (cf. Ref. \cite{JP95}). 
 

\section{Parameterization of the Dipole Fit}

Because no reliable self-consistent values of the Zemach correction are known
numerical results can be obtained only after reconsideration this correction. 
The parameter $\Lambda$ which is needed for this calcilation is 
directly connected with the proton square charge radius

\[
r_p = \frac{\sqrt{12}}{\Lambda}.
\]

Comparision of recent experimental results with the theory (see e. g. Ref. 
\cite{Pac96}) favors the newer value \cite{Sim}

\[
r_p = 0.862(12)~fm, 
\]

\noindent or

\beq
\Lambda = 0.845(12) ~m_p.
\eeq

\noindent 
However, as most of the work in this field 
was done more than 15 years ago,
the older proton radius of Ref. \cite{Han}

\[
r_p = 0.809(11)~fm, 
\]

\noindent or

\beq
\Lambda = 0.898(13) ~m_p
\eeq

\noindent
was used. It is well-known that the form factor from the older radius and the
value
$\Lambda = 0.898(13) ~m_p$ is  good as long as the momentum transfer is not 
too low.
Hence, the first problem is  to understand 
what momenta are important in the integration of \eq{ZemEx}. In order 
to solve this problem high and low momenta have been separeted in the 
integral

\beq \label{Zem-Q}
\Delta \nu(Zemach) = \nu_F \, \frac{8Z\alpha}{\pi}\frac{m_e}{m_p}
\times
\]

\[
\left\{
m_p\int_{0}^{Q}{\frac{dp}{\bf p^2}}
\left[ \left(G_D({\bf p^2})\right)^2-1 \right]+
m_p\int_{Q}^{\infty}{\frac{dp}{\bf p^2}}
\left[ \left(G_D({\bf p^2})\right)^2-1\right]
\right\}
.\eeq

\noindent
The low momenta asymptotics has the form

\beq  \label{LAs}
\left[ \left(G_D({\bf p^2})\right)^2-1 \right] \simeq
-4\,\frac{{\bf p}^2}{\Lambda^2}
,\eeq

\noindent
and the  high momenta asymptotic behaviour is

\beq \label{HAs}
\left[ \left(G_D({\bf p^2})\right)^2-1\right] \simeq -1
.\eeq

The results of integrations are presented in Fig.1.


The main radius-dependent contribution arises from 
the low momenta asymptotics ($-4{\bf p^2}/\Lambda^2$), which should be directly 
connected to the radius. Our results for several $Q$ (see also Fig. 2)

\beq  \label{QZem}
\Delta \nu(Zemach) =
\left\{
\begin{array}{cc}
-40.92(59)\cdot 10^{-6} \cdot \nu_F,~~&~~Q=0.30~m_p,\\
&\\
-41.07(68)\cdot 10^{-6} \cdot \nu_F,~~&~~Q=0.35~m_p,\\
&\\
-41.24(68)\cdot 10^{-6} \cdot \nu_F,~~&~~Q=0.40~m_p,\\
\end{array}
\right.
\eeq

\noindent
are obtained as the sum 
of the $4{\bf p^2}/\Lambda^2$-contibution from the lower momenta with 
$\Lambda=0.845\, m_p$ and the average value 
of the remaining terms with 
$\Lambda=0.845\, m_p$ and $\Lambda=0.898\, m_p$. The uncertainty is obtained from
the sum of the squares of the parameter-induced uncertainty in the 
$4{\bf p^2}/\Lambda^2$ term 
and half of the difference for the contribution of
$G_D^2-1-4{\bf p^2}/\Lambda^2$ and of higher momenta
with $\Lambda=0.845\, m_p$ and $\Lambda=0.898\, m_p$. The details of 
calculations are contained in the appendix.

The results for different values of $Q$ are 
almost independent of $Q$. They should be compared with results 
for the point-like proton

\beq  \label{rZem}
\Delta \nu(Zemach) =
\left\{
\begin{array}{cc}
-41.15(58)\cdot 10^{-6} \cdot \nu_F,&\Lambda=0.845(12)\, m_p,\\
&r_p=0.862(12)\,fm,\\
&\\
-38.72(56)\cdot 10^{-6} \cdot \nu_F,&\Lambda=0.898(13)\, m_p, \\
&r_p=0.809(11)\,fm,\\
&\\
-40.43(43)\cdot 10^{-6} \cdot \nu_F,&\Lambda=0.860(9)\, m_p, \\
&r_p=0.847(9)\,fm.\\
\end{array}
\right.
\eeq

\noindent
The last proton radius value is the result 
presented in the recent work \cite{newestR}.

One can see that the results of our estimate in \eq{QZem} are close to 
the result of \eq{rZem} for $\Lambda=0.845(12)\, m_p$, but the
uncertainties are a little higher. This agreement of \eq{QZem} with \eq{rZem} 
for $\Lambda=0.845(12)\,
m_p$ is due to the use of this value in our calculation of
the $4{\bf p^2}/\Lambda^2$ term. 
Within the interval between $0.3\,m_p$ and $0.4\,m_p$ 
of value of Q  the approach described here is expected to yields the best results.
The value 

\beq  \label{fZem}
\Delta \nu(Zemach) =
-41.07(75)\cdot 10^{-6} \cdot \nu_F
\eeq

\noindent
is used
for futher numerical calculation. We expect that this result is more safe
than any simple dipole fit  values from \eq{rZem}.


\section{Conclusion}

In one of the latest works \cite{Bod}, devoted to 
the ground state hyperfine splitting in  the hydrogen atom, compararison 
of theory and experiment leads to the difference

\beq \label{BoComp}
\frac{\nu_{HFS}(exp)-\nu_{HFS}(theo)}{\nu_{HFS}(exp)}=
\Big(0.56  \pm 0.48\Big)\,ppm.
\eeq
 
\noindent
The theoretical exression exludes the unknown proton polarizability so it may
be  estimated by the difference in \eq{BoComp}. The theoretical limitation
for the proton polarizability contribution is \cite{pol1,pol2,pol3}

\beq  \label{Lim}
\vert\delta(polarizability)\vert<4\,ppm.
\eeq

The result of this work is 

\beq \label{MyComp}
\frac{\nu_{HFS}(exp)-\nu_{HFS}(theo)}{\nu_{HFS}(exp)}=
\Big(3.5  \pm 0.9\Big)\,ppm
\eeq
 
\noindent
instead \eq{BoComp}. The changes of the theoretical values are presented in
Tables 1--3. The older comparission of theory and experiment in \eq {BoComp}
implies  that the polarizability contribution is much lower than the 
limitation of 
\eq{Lim}. However, our comparision  in \eq {MyComp} leads to a result close
to this limitation.

\bigskip

The hyperfine splitting of the hydrogen ground state is more sensitive 
to the proton structure value than the Lamb shift. We hope that 
investigation of the hfs will lead to a better understanding of the proton 
and to a more accurate calculation of the Lamb shift, the precision of which
is limitted by the proton radius.

\bigskip

The author is grateful to K. Jungmann for stimulating discussions. 
The final part of this work was done 
durinig the author«s summer stay at the Max-Planck-Institut f\"ur 
Quantenoptik 
and he 
is very grateful to T. W. H\"ansch for his hospitality.  The author would 
like to 
thank Thomas Udem for reading the manuscript and useful remarks.

This work was supported in part by the grant
\#95-02-03977 of the Russian Foundation for Basic Research.
The presentation of the work at the ZICAP is supported by the Organizing
Commitee and the author would like to thank them for their support.

\newpage
 
\appendix
\section{Q-integrals for calculation of the Zemach correction}

1. The integrals over different area are 

\[
I_<(Q,\Lambda)=m_p\int_{0}^{Q}{\frac{dp}{\bf p^2}}
\left[ \left(G_D({\bf p^2})\right)^2-1 \right]
\]

\[
=
-\frac{35}{16}\frac{m_p}{\Lambda}\,arctg\left(\frac{Q}{\Lambda}\right)
-\frac{19}{16}\frac{m_p Q}{\Lambda^2+Q^2}
-\frac{11}{24}\frac{m_p Q \Lambda^2}{\left(\Lambda^2+Q^2\right)^2}
-\frac{1}{6}\frac{m_p Q \Lambda^4}{\left(\Lambda^2+Q^2\right)^3}
\]

\noindent and 

\[
I_>(Q,\Lambda)=m_p
\int_{Q}^{\infty}{\frac{dp}{\bf p^2}}
\left[ \left(G_D({\bf p^2})\right)^2-1\right]
\]

\[
=
-\frac{35}{16}\frac{m_p}{\Lambda}\,arctg\left(\frac{\Lambda}{Q}\right)
+\frac{19}{16}\frac{m_p Q}{\Lambda^2+Q^2}
+\frac{11}{24}\frac{m_p Q \Lambda^2}{\left(\Lambda^2+Q^2\right)^2}
+\frac{1}{6}\frac{m_p Q \Lambda^4}{\left(\Lambda^2+Q^2\right)^3}.
\]
 
\noindent
2. The contributions of asymptotics of \eq{LAs} and \eq{HAs} are easy to
find:

\[
A_<(Q,\Lambda) = -4\, \frac{m_pQ}{\Lambda^2}
\]

\noindent
and 

\[
A_>(Q,\Lambda) = - \frac{m_p}{Q}.
\]

\noindent
The contributions of remaining low-momenta term is denoted by

\[
R_<(Q,\Lambda)=I_<(Q,\Lambda)-A_<(Q,\Lambda) 
.\]

\noindent
{\sc All results with different 
$\Lambda$ are presented in Fig. 1 as functions of $Q$.}\\

\noindent
3. The following combinations are used as the result and the
uncertainty

\[
I(Q)=A_<(Q,0.845\,m_p) +\frac{R_<(Q,0.845\,m_p)+R_<(Q,0.898\,m_p)}{2}
\]

\[
+
 \frac{I_>(Q,0.845\,m_p)+I_>(Q,0.898\,m_p)}{2}
\]

\noindent and 

\[
\delta I(Q)=\Bigg\{  \left(\frac{8\,\delta\, \Lambda\,m_p\,Q}
{\Lambda^3}\right)^2
\]

\[
+
\left(\frac{R_<(Q,0.845\,m_p)-R_<(Q,0.898\,m_p)}{2}\right)^2+
\left(\frac{I_>(Q,0.845\,m_p)-I_>(Q,0.898\,m_p)}{2}\right)^2
\Bigg\}^{1/2}.
\]

\noindent The function $I(Q)$ is presented in Fig. 2.

\newpage

\section*{List of captions to the tables and the figures}

Table 1: Old and new values of some relative contributions to the theoretical
hfs (in ppm).\\ 
${}^a$ The following new values have been used: $\alpha^{-1}=137.0359994(6)$
from Ref. \cite{alp} and
$m_p/m_e=1836.152667(4)$ from Ref. \cite{mp}.\\ 
${}^b$ The result of Refs. \cite{Brod,Sap832} used in 
Ref. \cite{Bod} is incorrect (for details see review \cite{ZPd} and Refs. 
\cite{Sch,Pri}).\\
${}^c$ This is actually a result of Ref. \cite{Bod}, but the older proton
radius was used there.
The uncertainty given there was tripled  here.\\

\noindent
Table 2: Contributions to the relative shift of the theoretical result for 
the hfs (in  ppm).\\

\noindent
Table 3: Contributions to the uncertainty of the theoretical result for the 
hfs (in ppm).\\

\noindent
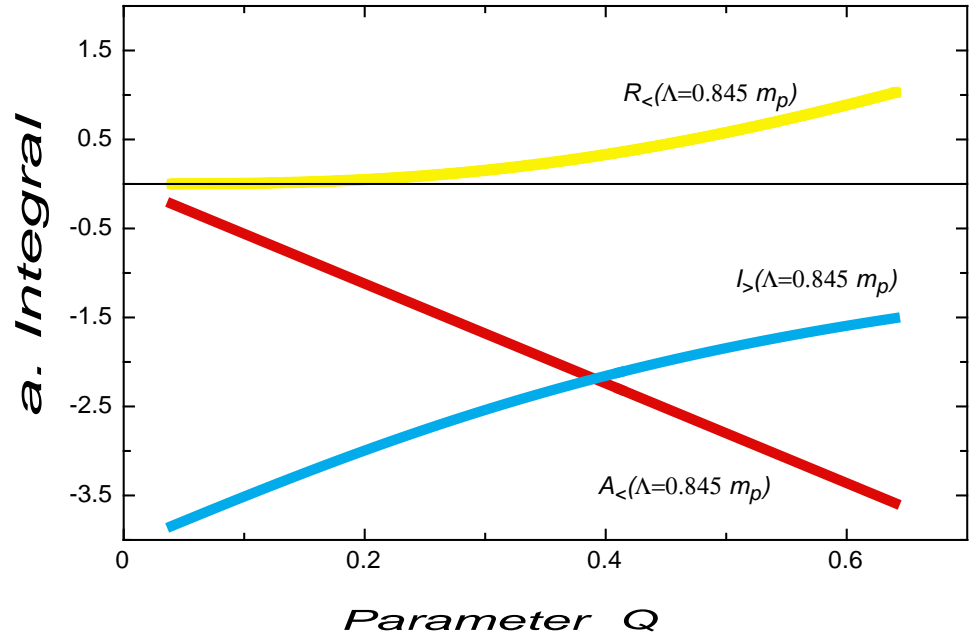
Fig 1: Contribution to the integral in \eq{Zem-Q} for 
$\Lambda=0.898\, m_p$. $I_>$ -- contribution from higher momenta,  $A_<$ --
contribution of asymptotics of \eq{LAs} from lower momenta and $R_<$ 
contribution of the remaining terms.\\

\noindent
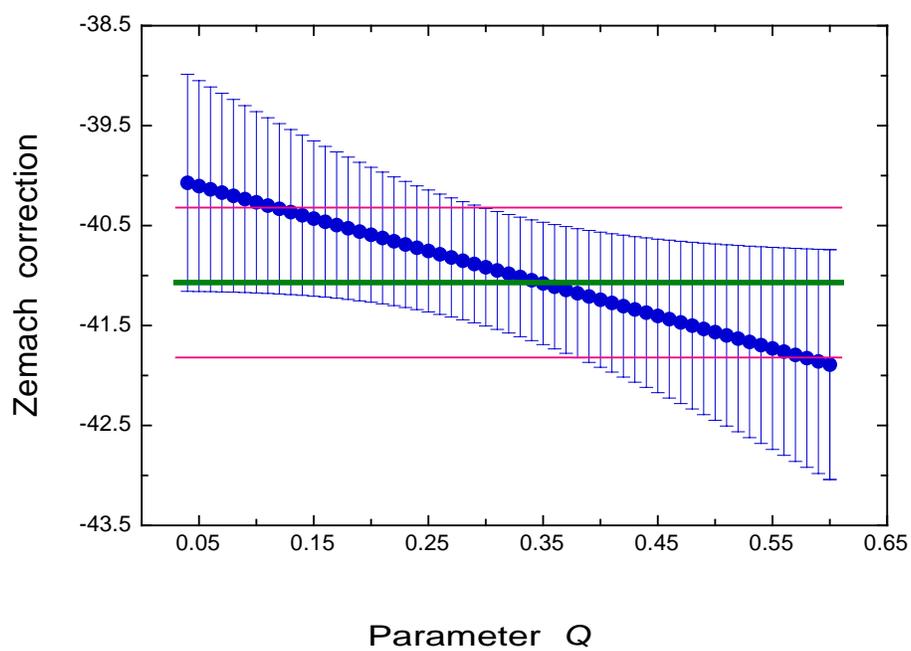
Fig 2: The Zemach correction as a function of $Q$.\\

\newpage

\begin{table}
\begin{center}
\vspace*{5mm}
\begin{tabular}{||c|c|c|c||}
\hline
\hline
&&&\\[-1ex]
Term & Old & New & Ref. \\[1ex]  
\hline
&&&\\[-1ex]
$\nu_F-\nu_{exp}$  & -1102.15 & -1102.28 & ${}^a$\\ [1ex] 
$\alpha(Z\alpha)^2$  &  1.90(4) & 1.84(12) & ${}^b$\\ [1ex] 
$\alpha^2(Z\alpha)$  & --  & 0.09 &\cite{IEEE,KN,ES}\\ [1ex] 
$higher\; ord.$  & --  & 0.01(2) &\cite{JETP93,ZPd}\\ [1ex] 
$VP-Structure$  & --  & -0.74(1) & this work\\ [1ex] 
$SE-Structure$  & --  &  0.12 & this work \\ [1ex] 
$Binding-Structure$  &  & -0.01& this work\\ [1ex] 
$Zemach$  &  -38.72(56) &-41.07(75) & this work\\ [1ex]
$Recoil-Structure$  & 5.22(14) & 5.22(42)& ${}^c$\\ [1ex] 
$Recoil-VP-Structure$  & --  & -0.02 & this work\\ [1ex] 
$Recoil-SE-Structure$  & --  & 0.11(2) & this work\\ [1ex] 
$\mu-VP-Structure$  & --  & 0.07(2) & this work\\ [1ex] 
$Hadr-VP-Structure$  & --  & 0.03(1) & this work\\ [1ex]
$Weak \;\; interaction$  & --  & -0.06 & \cite{weak}\\ [1ex]
\hline
\hline
\end{tabular}
\end{center}
\caption{
}
\end{table}

\begin{table}
\begin{center}
\vspace*{5mm}
\begin{tabular}{||c|c||}
\hline
\hline
&\\[-1ex]
Term & Shift   \\[1ex]  
\hline
&\\[-1ex]
$Constants\; (\alpha\;,m_e/m_p)$  & -0.13  \\ [1ex] 
$QED$  &  0.04   \\ [1ex] 
$Zemach$  & -2.36   \\ [1ex] 
$Radiative-Structure$   &  -0.63 \\ [1ex] 
$RRC-Structure$ &  0.09  \\ [1ex] 
$Heavy-VP-Structure$   & 0.10\\ [1ex]
$Weak \;\; interaction$  & -0.06 \\ [1ex]
\hline
&\\[-1ex]
$Total$  &  -2.95  \\ [1ex] 
\hline
\hline
\end{tabular}
\end{center}
\caption{}
\end{table}

\begin{table}
\begin{center}
\vspace*{5mm}
\begin{tabular}{||c|c||}
\hline
\hline
&\\[-1ex]
Term & Uncertainty \\[1ex]  
\hline
&\\[-1ex]
$QED$  & 0.12   \\ [1ex] 
$Zemach$  &  0.75   \\ [1ex] 
$Radiative-Structure$     & 0.03 \\ [1ex] 
$Recoil$  &   0.42 \\ [1ex] 
$RRC-Structure$     & 0.03 \\ [1ex] 
$Heavy-VP-Structure$   & 0.02\\ [1ex]
\hline
&\\[-1ex]
$Total$  & 0.87   \\ [1ex] 
\hline
\hline
\end{tabular}
\end{center}
\caption{}
\end{table}


\newpage

\begin{thebibliography}{99}

\bibitem{HFS1} H. Hellwig, R. F. C. Vessot et al., 
IEEE Trans. Instr. and Meas. {\bf IM 19} (1970) 200.

\bibitem{HFS2} L. Essen, R. W. Donaldson et al., Nature (London) {\bf 229} (1971)
 110. 

\bibitem{Zem} A. C. Zemach, Phys. Rev. {\bf 104} (1957) 1771.

\bibitem{Arn} R. Arnowitt, Phys. Rev. {\bf 92} (1953) 1002.

\bibitem{New} W. A. Newcomb and E. E. Salpeter,  Phys.  Rev. {\bf 97}
(1955) 1146.

\bibitem{Idd} C. K. Iddings and P. M. Platzman, Phys.  Rev. {\bf 113}
(1959) 192.

\bibitem{Gro} H. Grotch and D. R. Yennie, Z. Phys. {\bf 202} (1967) 425;
Rev. Mod. Phys. {\bf 41} (1969) 350.

\bibitem{Bod} G. T. Bodwin and D. R. Yennie, Phys. Rev. {\bf D37} (1988)
498.

\bibitem{KSE} S. G. Karshenboim, V. A. Shelyuto and M. I. Eides, Yad.
Fiz. {\bf 50} (1989) 1636 /in Russian/; Sov. J. Nucl. Phys. {\bf 50}
(1989) 1015.

\bibitem{STY} J. R. Sapirstein, E. A. Terray, and D. R. Yennie, Phys.
Rev. {\bf D29} (1984) 2990.

\bibitem{EKS91} M. I. Eides, S. G. Karshenboim and V. A. Shelyuto, Ann.
Phys. {\bf 205} (1991) 231.

\bibitem{BEKS} V. Yu. Brook, M. I. Eides, S. G. Karshenboim and V. A.
Shelyuto, Phys. Lett. {\bf 216B} (1989) 401.

\bibitem{JETP93} S. G. Karshenboim, ZhETF {\bf 103} (1993) 1105
/in Russian/; JETP {\bf 76} (1993) 541.

\bibitem{IK96NP} V. G. Ivanov and S. G. Karshenboim,
{\sl 1996 Conference on Precision Electromagnetic Measurements.\/} Conference 
digest, 636. Braunschweig, 1996; Yad. Fiz. (1996) /in Russian/;
Phys. At. Nucl. (1996), to be published.

\bibitem{log2} S.~G.~Karshenboim, 
ZhETF {\bf 106} (1994) 414 /in Russian/; JETP {\bf 79} (1994) 230;\\
Yad. Fiz. {\bf 58} (1995) 707 /in Russian/; Phys. At. Nucl. {\bf 58} (1995) 649;\\
ZhETF {\bf 109} (1996) 752 /in Russian/; JETP {\bf 82} (1996) 403;\\
J.~Phys. {\bf B29} (1996) L21.

\bibitem{JP95} S. G. Karshenboim, J. Phys. {\bf B 28} (1995) L77.

\bibitem{Pac96}  M.~Weitz, A.~Huber A.~F. Schmidt-Kaler, D.~Leibfried,
W.~Vassen, C.~Zimmermann, K.~Pachucki, T.~W.~H\"ansch, L.~Julien and
F.~Biraben,
Phys. Rev.  {\bf 75} (1995) 2664;\\ 
K. Pachucki, D. Leibfried, M. Weitz, A. Huber, W. K\"onig
and T. W. H\"ansch, J.~Phys. {\bf B29} (1996) 177; {\sl ibid}., 1573.

\bibitem{Sim} G. G. Simon,  Ch. Schmitt et al. Nucl. Phys. {\bf A333}
(1980) 381.

\bibitem{Yen} D. R. Yennie, Zeitschrift f\"ur Physik {\bf C36} (1992) S13.

\bibitem{Han} L. Hand, D. I. Miller and R. Willson, Rev. Mod. Phys. {\bf
35} (1963) 335.

\bibitem{newestR} P. Mergell, U.G. Meissner and D. Drechsel, Nucl. Phys.  
{\bf A596} (1996) 367.

\bibitem{pol1} V. W. Hughes and J. Kuti, Annu. Rev. Nucl. Part. Sci. {\bf 33} (1983)
611.

\bibitem{pol2} E. de Rafael, Phys. Lett. {\bf 37B} (1971) 201.

\bibitem{pol3}  P. Gn\"adig and J. Kuti, Phys. Lett. {\bf 42B} (1972) 241.

\bibitem{IEEE}  M. I. Eides, S. G. Karshenboim and V. A. Shelyuto,
IEEE Trans. Instr. and Meas. {\bf IM 44} (1995)
481.

\bibitem{KN} T. Kinoshita and M. Nio, Phys. Rev. Lett. {\bf 72} (1994)
3803.

\bibitem{ES} M. I. Eides and V. A. Shelyuto, Phys. Rev. {\bf A52} (1995)
954.

\bibitem{JETP93} S. G. Karshenboim, ZhETF {\bf 103} (1993) 1105
/in Russian/; JETP {\bf 76} (1993) 541.

\bibitem{ZPd} S. G. Karshenboim, Zeitschrift f\"ur Physik {\bf D36} (1996) 11.

\bibitem{weak} M. A. B. B\'eg and G. Feinberg, Phys. Rev. Lett. {\bf 33} (1974)  
606;  {\sl ibid}., {\bf 35} 130.


\bibitem{alp} T. Kinoshita, Phys. Rev. Lett. {\bf 75} (1995) 4728.

\bibitem{mp} D. L. Farnham, R. S. Van Dyck, Jr., and P. B. Schwinberg, 
Phys. Rev. Lett. {\bf 75} (1995) 3598.

\bibitem{Brod} S. J. Brodsky and G. W. Erickson, Phys. Rev. {\bf 148}
(1966) 26.

\bibitem{Sap832} J. R. Sapirstein, Phys. Rev. Lett. {\bf 51} (1983) 985.

\bibitem{Sch} S. M. Schneider, W. Greiner and G. Soff, Phys. Rev. A {\bf
50} (1994) 118.

\bibitem{Pri}  It was also discovered by M. Nio and
by J.~R.~Sapirstein, independently, and confirmed by S. J. Brodsky and G.
W. Erickson. The author is very grateful to T. Kinoshita and M. Nio for private
communications on that.



\end{thebibliography}
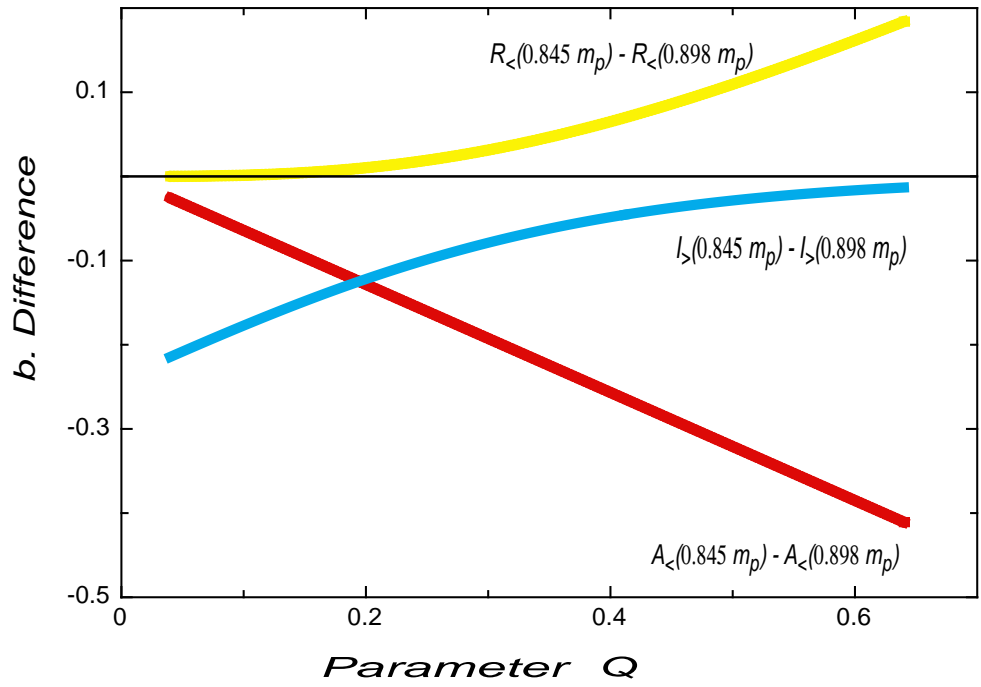
\end{document}